\documentclass{article}
\usepackage{lmrl2025_workshop,times}
\usepackage{hyperref}
\usepackage{url}
\usepackage{graphicx}
\usepackage{amsmath}
\usepackage{amssymb}
\usepackage{booktabs}
\usepackage{caption}
\usepackage{subcaption}
\usepackage{array}
\usepackage{float}
\usepackage{microtype}

\usepackage{amsmath,amsfonts,bm}

\def\eqref#1{equation~\ref{#1}}

\def\1{\bm{1}}

\DeclareMathAlphabet{\mathsfit}{\encodingdefault}{\sfdefault}{m}{sl}
\SetMathAlphabet{\mathsfit}{bold}{\encodingdefault}{\sfdefault}{bx}{n}

\title{Leveraging State Space Models in Long Range Genomics}

\author{
    \hspace{1cm} \textbf{Matvei Popov}\thanks{equal contribution} \quad
    \textbf{Aymen Kallala}\footnotemark[1] \quad
    \textbf{Anirudha Ramesh}\footnotemark[1]~~\thanks{corresponding author - alternate email: anirudha.ramesh50@gmail.com} \quad
    \textbf{Narimane Hennouni} \\ \\ 
    \hspace{3cm}\textbf{Shivesh Khaitan} \quad \textbf{Rick Gentry} \quad \textbf{Alain-Sam Cohen} \\ \\ 
    \hspace{6cm}\textbf{InstaDeep} \\ \\ 
    \hspace{4.5cm}\texttt{a.ramesh@instadeep.com}
}

\lmrlfinalcopy 
\begin{document}

\maketitle

\begin{abstract}
Long-range dependencies are critical for understanding genomic structure and function, yet most conventional methods struggle with them. Widely adopted transformer-based models, while excelling at short-context tasks, are limited by the attention module's quadratic computational complexity and inability to extrapolate to sequences longer than those seen in training. In this work, we explore State-Space Models (SSMs) as a promising alternative by benchmarking two SSM inspired architectures, Caduceus and Hawk, on long-range genomics modeling tasks under conditions parallel to a 50M-parameter transformer baseline. We discover that SSMs match transformer performance and exhibit impressive zero-shot extrapolation across multiple tasks, handling contexts 10–100$\times$ longer than those seen during training, indicating more generalizable representations better suited for modeling the long and complex human genome. Moreover, we demonstrate that these models can efficiently process sequences of 1M tokens on a single GPU, allowing for modeling entire genomic regions at once, even in labs with limited compute. Our findings establish SSMs as efficient and scalable for long-context genomic analysis.
\end{abstract}

\maketitle

\begin{abstract}
Long-range dependencies are critical for understanding genomic structure and function, yet most conventional methods struggle with them. Widely adopted transformer-based models, while excelling at short-context tasks, are limited by the attention module's quadratic computational complexity and inability to extrapolate to sequences longer than those seen in training. In this work, we explore State-Space Models (SSMs) as a promising alternative by benchmarking two SSM inspired architectures, Caduceus and Hawk, on long-range genomics modeling tasks under conditions parallel to a 50M-parameter transformer baseline. We discover that SSMs match transformer performance and exhibit impressive zero-shot extrapolation across multiple tasks, handling contexts 10–100$\times$ longer than those seen during training, indicating more generalizable representations better suited for modeling the long and complex human genome. Moreover, we demonstrate that these models can efficiently process sequences of 1M tokens on a single GPU, allowing for modeling entire genomic regions at once, even in labs with limited compute. Our findings establish SSMs as efficient and scalable for long-context genomic analysis.

\end{abstract}

\section{Introduction}

Genomes are the fundamental blueprint of life. Advances in DNA sequencing have rapidly lowered costs, enabling the curation of high-quality genomic datasets and opening new avenues to understand complex biological processes. Yet, a critical challenge remains in modeling the long-range interactions inherent to genomic data, which can span billions of base pairs (e.g. $\approx$3 billion in the human genome). 

A single human chromosome can span hundreds of millions of nucleotides, with regulatory elements often residing hundreds of kilobases or more from their target genes. Subtle variations, such as single-nucleotide polymorphisms (SNPs), can disrupt these regulatory landscapes by changing enhancer or promoter activity, sometimes resulting in substantial phenotypic effects. As these elements and variations are interspersed throughout massive stretches of DNA, any method that cannot maintain full sequence context while also distinguishing base-level changes risks missing critical genomic signals. Models that rely on chunking sequences into shorter windows, or that tokenize DNA at the k-mer level, frequently lose the global view needed to understand how distant elements interact. Capturing ultralong (in this paper ultralong refers to 1Mbp+ sequences) context at single-base resolution is therefore central to revealing the intricate networks that govern gene regulation, disease susceptibility, and evolutionary adaptation. At the same time, the long context necessitates efficient training with respect to memory and computational complexity. Being able to extrapolate beyond sequence lengths seen during training proves to be a decisive advantage in the same, as it reduces the training burden without sacrificing the ability to evaluate the longest sequences.

Advances in machine learning have begun to transform genomics. Protein language models now reliably predict the effects of coding mutations on protein function \citep{doi:10.1126/science.ade2574}, generate viable protein sequences conditioned on functional properties \citep{madani2023large}, and accurately predict protein structures \citep{doi:10.1126/science.ade2574}, aided by benchmarks such as CASP \citep{kryshtafovych2021critical}, TAPE \citep{rao2019evaluating}, PEER \citep{xu2022peer}, and ProteinGym \citep{notin2024proteingym}. Following these advances, DNA-focused language models---including Nucleotide Transformer \citep{dalla2023nucleotide}, DNABERT \citep{DNABert}, and GPN-MSA \citep{Benegas2025GPNMSA}---have emerged to assist in tasks like motif discovery and gene annotation. However, while transformer-based models \citep{vaswani2023attentionneed} excel in short-context settings, their computational demands and quadratic scaling with sequence length limit their ability to capture truly long-range dependencies.

Recently, State-Space Models (SSMs) \citep{gu2022efficientlymodelinglongsequences} have shown promise as an efficient alternative. Their linear complexity with respect to sequence length allows them to maintain rich contextual information over extended spans. Here, we evaluate two classes of SSM-inspired architectures ---Caduceus \citep{schiff2024caduceusbidirectionalequivariantlongrange} and Hawk \citep{de2024griffinmixinggatedlinear}, which are built on top of Mamba \citep{gu2024mambalineartimesequencemodeling} and LRU \citep{orvieto2023resurrectingrecurrentneuralnetworks}, on DNA modeling tasks. We find that SSMs not only match transformer performance under standard settings but also demonstrate a striking capacity for zero-shot extrapolation to sequence lengths up to 10-100$\times$ beyond those encountered during training, with trends indicating this can be pushed even further. This ability makes them strong candidates for handling ultralong context lengths, a property that is invaluable for tackling the complexities of genomic sequences. We demonstrate this on multiple different genomics tasks, indicating rich and meaningful underlying representations. We also show that SSMs can scale to process sequences of 1Mbp+ on a single GPU, further underscoring their potential for large-scale genomic analysis.

Our specific findings are:

\begin{itemize}
    \item SSM-based models achieve performance on par with attention-based models on a wide range of DNA modeling tasks.
    \item SSMs zero-shot extrapolate to much longer contexts (10-100x) without additional finetuning, suffering minimal performance loss, with trends suggesting possible extrapolation to even longer contexts.
    \item We demonstrate scalability to 1Mbp+ sequences at single nucleotide-level on just one GPU, laying the groundwork for future large-scale genomic modeling.
\end{itemize}

\section{Background and Related Works} \label{sec:background}

\subsection{Long-Range Genomics Modeling and Benchmarks} \label{sec:long-range}

Understanding genomic functions often requires modeling interactions across vast distances, as regulatory elements like enhancers can influence gene expression over hundreds of kilobases. Traditional benchmarks focusing on short sequences fail to capture these extensive dependencies, limiting progress in long-range genomic modeling. The \textbf{Genomics Long-Range Benchmark (GLRB)}~\citep{kao2024advancing, InstadeepLtd} fills this gap by providing a suite of tasks—variant effect prediction, gene expression prediction, regulatory element detection, and chromatin feature identification—with inputs ranging from a few to hundreds of kilobases. By standardizing evaluation on both short- and long-range contexts, GLRB not only catalyzes advances in modeling extensive genomic interactions but also serves as a biologically meaningful indicator of a model's ability to capture the long-range dependencies underlying enhancer-promoter interactions, chromatin remodeling, and gene regulation.

\subsection{Limitations of Existing Models in Long-Range Genomics} \label{sec:current-approaches}

While transformer-based models have advanced genomic sequence modeling, they encounter significant limitations when applied to ultralong sequences due to computational constraints.

Nucleotide Transformer \citep{dalla2023nucleotide}  (NT): NT employs a transformer architecture with masked language modeling to learn dependencies within DNA sequences, achieving strong performance in tasks like splice site detection and enhancer classification. Recent efforts have led to extend the context length of pretrained NT models using extrapolation techniques such as YaRN\citep{peng2023yarnefficientcontextwindow}, however these still require a fine-tuning with new sequence lengths, and don't work zero-shot.

DNABERT~\citep{DNABert}: DNABERT adapts the BERT architecture for genomic data using k-mer tokenization to process DNA sequences. It excels in motif detection and regulatory sequence prediction but faces limitations in handling longer sequences due to its fixed tokenization scheme and strict sequence length limit (e.g., 512 tokens), hindering its ability to capture long-range interactions.

HyenaDNA~\citep{nguyen2023hyenadnalongrangegenomicsequence}: HyenaDNA represents a shift from transformer-based approaches by adopting implicit convolutional mechanisms to efficiently model long-range dependencies. It processes sequences up to 1 million tokens at the single nucleotide-level. 

Although HyenaDNA can process up to 1 million tokens—matching the 1Mbp performance reported for SSM in this paper—it significantly underperforms NTv2 on several GLRB tasks, including BulkRNA, CAGE, and Histone Marks Detection. \citep{kao2024advancing, InstadeepLtd}. This likely stems from its convolutional architecture, which lacks the representational capacity and interpretability of attention-based models. Therefore we use a better performing NTv2 as our primary comparison. 

\subsection{State Space Models for Efficient Long-Range Sequence Modeling}
\label{sec:ssm}

State Space Models (SSMs) have emerged as powerful architectures for long-range sequence modeling, particularly in genomics. Leveraging latent state representations and recurrent mechanisms, SSMs achieve linear complexity with respect to sequence length, making them well-suited for ultra-long genomic sequences (see Figure~\ref{fig:griffin_memory} in Appendix for more details). Recent advances in SSMs, such as time-dependent parameterization and bidirectional processing, have further enhanced their scalability and adaptability for dense genomic data~\citep{gu2022efficientlymodelinglongsequences, gu2024mambalineartimesequencemodeling, schiff2024caduceusbidirectionalequivariantlongrange, de2024griffinmixinggatedlinear}. While SSMs have shown promise in sequence modeling, they have not yet been systematically benchmarked for long-range genomic tasks, making our study one of the first to rigorously evaluate their scalability, zero-shot extrapolation, and practical feasibility in this domain.

\section{Experiments}
\label{headings}
Our experimental approach rigorously compares State-Space Models (SSMs) with state-of-the-art (SOTA) transformer-based architectures under a standardized evaluation framework, ensuring fair assessment. We begin by pre-training all models on the same dataset using similar model sizes, and then fine-tune them on Long-Range Genomics Benchmark (GLRB) tasks. As our baseline, we employ the NTv2 \citep{dalla2023nucleotide} transformer model, which represents SOTA in genomic tasks.

\subsection{Model Architectures}
We consider three classes of models, each with 50M parameters for a fair comparison: \\
\textbf{NTv2:} Our baseline is a smaller-variant Nucleotide Transformer model, which is the current SOTA model on GLRB.\\
\textbf{Caduceus:} Caduceus is the first successful application of an SSM to genomic tasks. \citep{schiff2024caduceusbidirectionalequivariantlongrange} It extends Mamba layers \citep{gu2024mambalineartimesequencemodeling} with bi-directionality and reverse-complement (RC) equivariance, showing promising genomics modeling results.\\
\textbf{Hawk:} Hawk is a recurrent architecture built on Linear Recurrence Units (LRUs) \citep{orvieto2023resurrectingrecurrentneuralnetworks}. It achieves competitive standard performance while excelling in zero-shot extrapolation to sequence lengths far beyond those seen during training. Inspired by Caduceus we enhanced Hawk with bi-directional processing.\\

\subsection{Pretraining}

For our experiments, we train our own version of Caduceus and Hawk, but use a pretrained version of NTv2. We follow the data sourcing and preprocessing procedures outlined in established genomic language model protocols \citep{dalla2023nucleotide, nguyen2023hyenadnalongrangegenomicsequence}. Specifically, we made use of the multispecies genome used to pretrain NTv2 in the exact same setting \citep{dalla2023nucleotide}, and pretrained our models on 300B nucleotides. Standard preprocessing steps are applied: tokenization treats each nucleotide (A, C, G, T, plus padding or masking tokens) as a single token to ensure full single-nucleotide resolution. We employ a masked language modeling (MLM) objective, randomly masking approximately 15\% of tokens in each input sequence, with sequences truncated or padded to a fixed length (e.g. 12 kbp) for batch consistency. The training protocol uses the AdamW \citep{loshchilov2019decoupledweightdecayregularization} optimizer with cosine learning rate scheduling. Although hyperparameters such as batch size, learning rate, and gradient clipping are kept consistent across model types, slight adjustments were made relative to the original NTv2 values since we found that mamba-based architectures tend to be more stable with larger batch sizes. Pretraining is carried out using data-parallel training across multiple GPUs.

\subsection{Fine-Tuning on Long-Range Genomics Tasks}

Following pretraining, the models are fine-tuned on a suite of long-range genomics tasks from the GLRB benchmark \citep{kao2024advancing, InstadeepLtd}, which assess the ability to predict genomic features (e.g., regulatory elements, gene expression, chromatin marks) from long-context sequences. For NTv2 and Caduceus, fine-tuning was performed on six key tasks from GLRB, including Bulk RNA (R$^2$), VEP eQTL (AUROC), VEP ClinVar (AUROC), Histone Marks (AUPRC), Promoters (AUPRC), and Enhancers (AUROC). In contrast, Hawk was selectively fine-tuned only on the VEP tasks (VEP eQTL and VEP ClinVar). This decision was driven by our goal of testing Hawk's zero-shot extrapolation capability to process sequences up to 1Mbp via hidden state expansion, as it allowed easier adaptation than Caduceus which would have required extensive modifications to its Mamba block architecture. 

Table~\ref{experiment-table} summarizes the performance of NTv2, Caduceus, and Hawk. Notably, Caduceus achieves very competitive results, outperforming NTv2 on several tasks (e.g., Bulk RNA, Histone Marks, and Promoters), with all models being trained in a similar fixed setting. This observation underscores the strength of the Caduceus and SSMs overall in capturing long-range genomic dependencies, and having meaningful internal representations for genomics aiding generalization across various tasks. 
Hawk underperforms relative to Caduceus, likely due to its architecture being less adapted to genomics modeling. However, its results remain significant, as we later use them to assess the extrapolation performance of SSMs with 1Mbp.

\begin{table}[H]
\caption{Long-Range Genomics Benchmark results, using 12kbp per sequence. Caducues achieves the best result in three out of six GLRB tasks.}
\label{experiment-table}
\begin{center}
\begin{tabular}{|c|c|c|c|}
\hline
\multicolumn{1}{|c|}{\bf Task} & 
\multicolumn{1}{|c|}{\bf NTv2} & 
\multicolumn{1}{|c|}{\bf Caduceus} & 
\multicolumn{1}{|c|}{\bf Hawk} \\
\hline
Bulk RNA (R2)           & 0.52  & \underline{\textbf{0.53}} & - \\
VEP eQTL (AUROC)        & \textbf{0.72} & 0.68  & 0.60  \\
VEP ClinVar (AUROC)     & \textbf{0.75} & \textbf{0.75} & 0.55 \\
Histone Marks (AUPRC)   & 0.34  & \underline{\textbf{0.52}} & - \\
Promoters (AUPRC)       & 0.75  & \underline{\textbf{0.77}} & - \\
Enhancers (AUROC)       & \textbf{0.78} & 0.75  & -  \\
\hline
\end{tabular}
\end{center}
\end{table}

\subsection{Zero-shot extrapolation}
\begin{figure}[!htbp]
    \centering
    \includegraphics[width=0.65\textwidth]{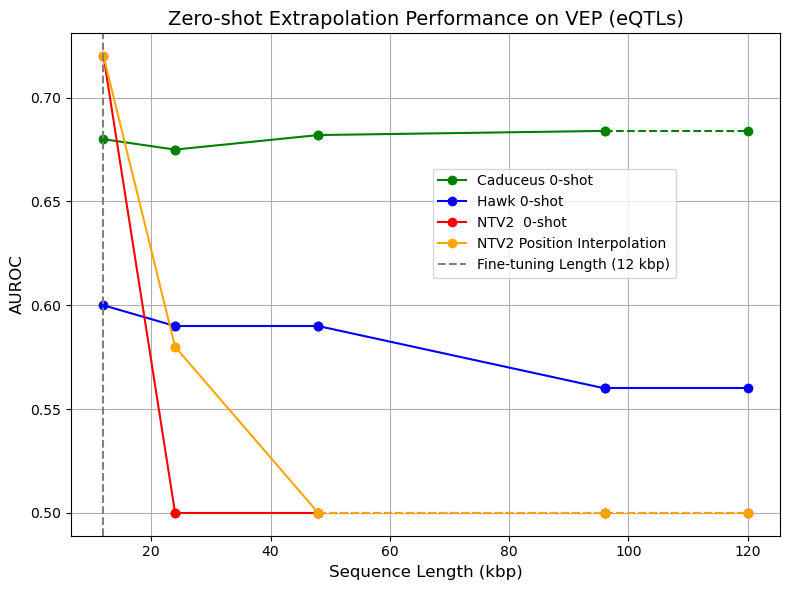}
    \caption{Comparison of the extrapolation methods of state-space models and attention-based models on VEP eQTLs (AUROC). For NTv2, we also reported an inference-time extrapolation method: position interpolation. A dotted vertical line indicates the fine-tuning sequence length (12 kbp) of all models. Attention-based models collapse when processing sequences that are longer than what they have encountered at training time, whereas state-space models show an ability to generalize to sequences up to 10x longer. Lines that turn into dotted indicate values that we were unable to compute due to computational cost constraints and are therefore assumed based on trends.}
    \label{fig:extrapolation_ntv2_50m}
\end{figure}

We evaluate models' ability to generalize to significantly longer sequences in a scalable way, without requiring further fine-tuning. Specifically, we assess zero-shot extrapolation by testing on downstream tasks input lengths up to 10× greater than those seen during pretraining. We observe that SSMs can perform well without additional changes, leveraging their ability to handle extended sequences. For transformer-based models, we experiment with inference-time position interpolation, shown to enhance performance. Performance is measured using metrics like AUPRC, AUROC, and $R^2$, on GLRB downstream tasks enabling a direct comparison of each architecture's generalization capability. Direct comparison of our equally sized and trained NTv2, NTv2 with Position Interpolation \citep{chen2023extending} within RoPE \citep{su2024roformer}, Caduceus, and Hawk models can be viewed in Figure~\ref{fig:extrapolation_ntv2_50m}, and extrapolation results on additional tasks for Caduceus are shown in Figure~\ref{fig:zero_shot_extrapolation}. 

\subsection{Performance Across Multiple Downstream Tasks}
\begin{figure}[H]
    \centering
    \includegraphics[width=0.3\textwidth]{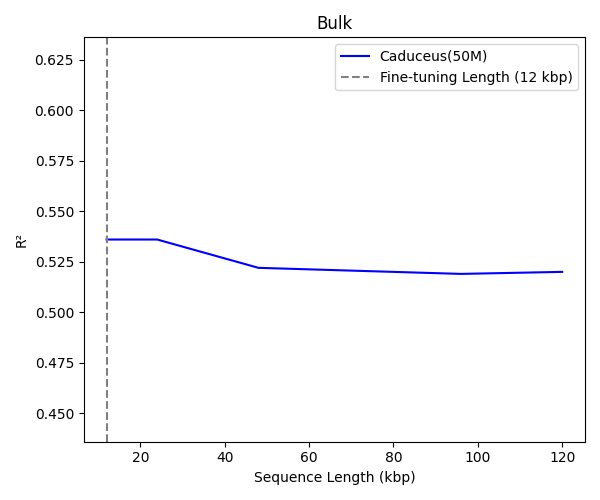}\hfill
    \includegraphics[width=0.3\textwidth]{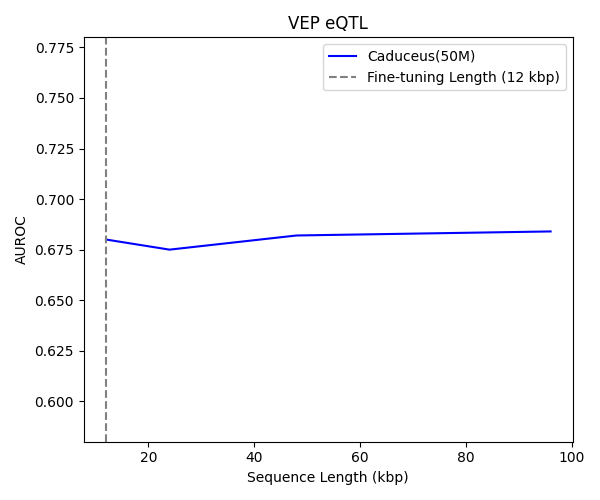} \hfill
    \includegraphics[width=0.3\textwidth]{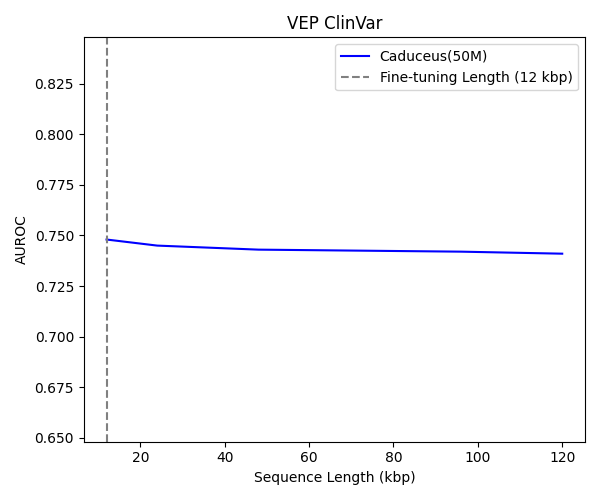} \\
    \includegraphics[width=0.3\textwidth]{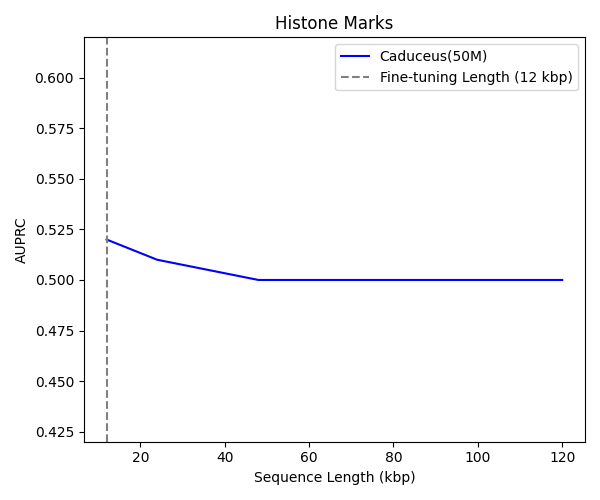} \hfill
    \includegraphics[width=0.3\textwidth]{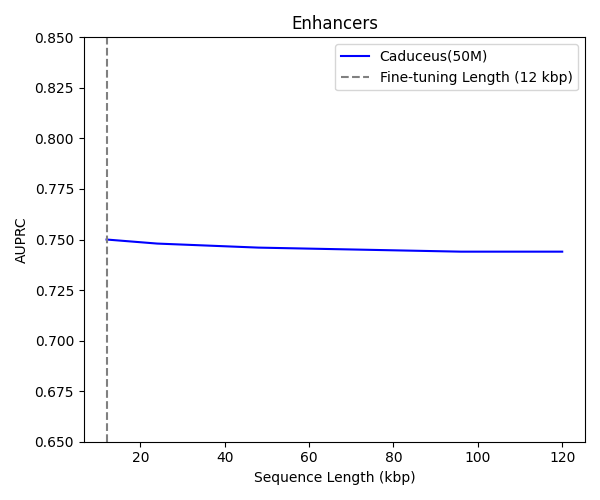} \hfill
    \includegraphics[width=0.3\textwidth]{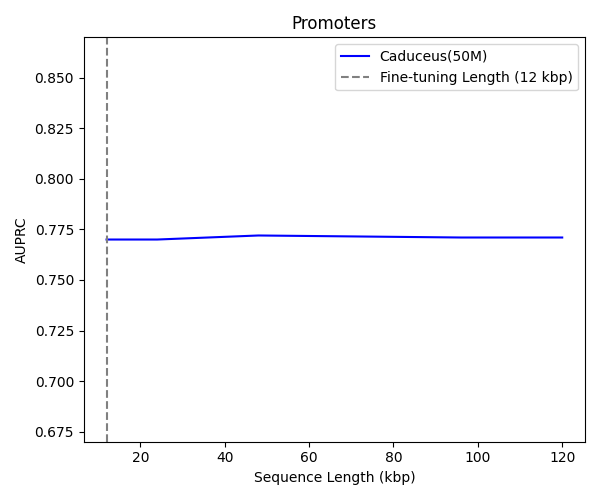} \\
    \caption{Zero-shot extrapolation results of state space models across the 6 tasks of the GLRB. Dotted vertical lines indicate the fine-tuning sequence length (12 kbp).}
    \label{fig:zero_shot_extrapolation}
\end{figure}

Caduceus achieves good zero-shot extrapolation results across all tasks (with sequence lengths ranging from 12\,kbp to 120\,kbp), as further visualized in Figure~\ref{fig:zero_shot_extrapolation}. These findings indicate that SSMs are not only capable of long-context extrapolation in a single task but also generalize effectively across diverse genomic prediction challenges, as well as their learned internal representations being very generalizable and high quality, allowing the model to confidently operate far beyond its training distribution.

\subsection{Processing Ultralong Sequences on a Single GPU}
\label{processing-ultralong-sequences}

\begin{figure}[H] \centering \includegraphics[width=0.7\textwidth]{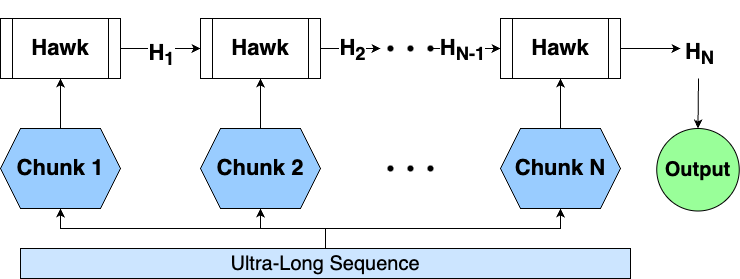} \caption{A Mechanism for Hidden State Propagation in SSMs for Ultralong Sequences Visualized.  An ultralong sequence is split into multiple chunks, thereby doing a linear scan over chunks. An individual chunk size could be set to any size that fits on a single GPU. The hidden state's size always stays fixed.} \label{fig:long_seq} \end{figure}

In this section, we demonstrate how hidden state transfer mechanism in SSMs can be used to process ultralong sequences of 1M+ tokens on a single GPU. As input sequences get longer, loading and processing it all at once, requires a large amount of memory. If an input sequence exceeds the maximum length that a single GPU can handle, the sequence is divided into smaller chunks (for example 100 kbp segments). The final hidden state from each chunk is passed as the initial state for the next chunk, ensuring continuity and preservation of dependencies across the entire sequence. This mechanism is intended to almost identically replicate a forward pass of the full ultralong sequence on a theoretical device with enough capacity (see Figure~\ref{fig:long_seq}).

\begin{figure}[!htbp]
    \centering
    \begin{subfigure}[b]{0.45\textwidth}
        \centering
        \includegraphics[width=\textwidth]{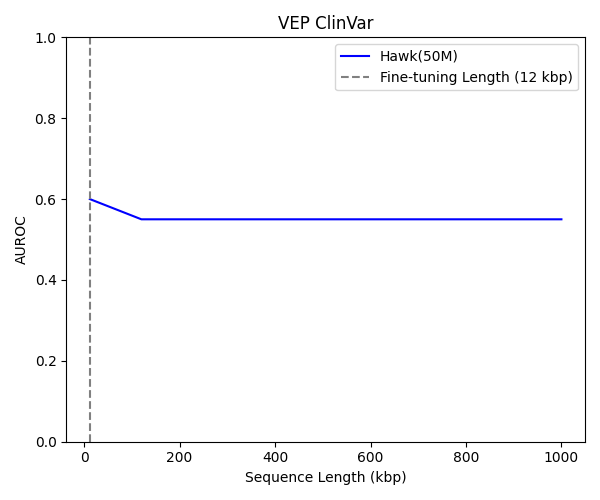}
        \label{fig:ClinVar}
    \end{subfigure}
    \hfill
    \begin{subfigure}[b]{0.45\textwidth}
        \centering
        \includegraphics[width=\textwidth]{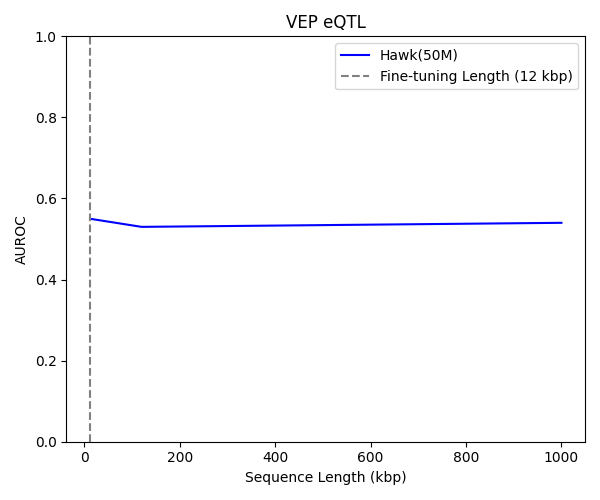}
        \label{fig:eQTL}
    \end{subfigure}
    \caption{Zero-shot extrapolation on VEP ClinVar and VEP eQTL with Hawk (50M parameters) up to 1 Mbp input length. Performance remains stable despite the substantial increase in context size, indicating strong scalability.}
    \label{fig:Hawk-1M}
\end{figure}

Building on our hidden state transfer mechanism for zero-shot extrapolation framework we adapt Hawk to process sequences up to 1Mbp on the VEP ClinVar and VEP eQTL tasks without any significant degradation in AUROC, on a single NVIDIA A100 GPU, highlighting its scalability and efficiency. 
A decision to use Hawk was made due to Hawk's simple mechanism for initializing and passing the hidden state as well as its simple implementation for linear scan. Future work can explore performing hidden state transfer on Caduceus where its custom kernel significantly complicates this task. Figure \ref{fig:Hawk-1M} illustrates Hawk's performance stability across the expanded sequences for the VEP ClinVar and VEP eQTL tasks. This marks a significant advancement of SSMs over standard transformer architectures. By scaling to 1Mbp dependencies, these models can capture long-range regulatory interactions that are often missed by traditional approaches. For instance, enable the detection of distal enhancers located hundreds of kilobases away from their target promoters, which is crucial for accurately linking non-coding variants to gene expression changes and disease phenotypes.

\subsection{Testing Extrapolation During Pretraining}
\begin{figure}[H]
    \centering
    \includegraphics[width=0.7\textwidth]{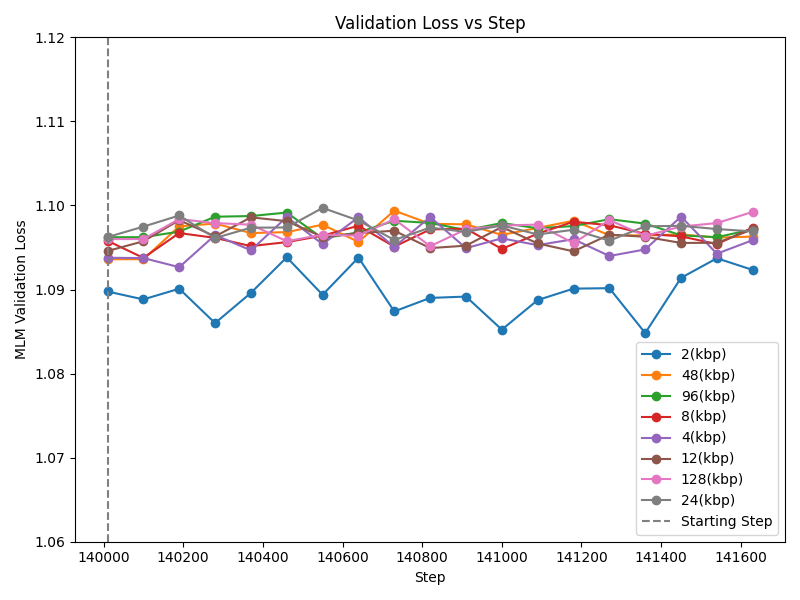}
    \caption{MLM validation loss as a function of training step for different sequence lengths (2 kbp, 4 kbp, 8 kbp, 12 kbp, 24 kbp, 48 kbp, 96 kbp, and 128 kbp). A dashed vertical line indicates the point at which training with 12 kbp sequences begins (the 140,000th training step). The y-axis shows the MLM loss, while the x-axis denotes training steps. Although the model continues training on a fixed 12 kbp context after this point, we measure validation loss across multiple lengths to assess generalization and extrapolation. The curves remain closely clustered, indicating that the model maintains comparable loss values even as the sequence length changes significantly.}
    \label{fig:val_loss_comp}
\end{figure}

We test if the nature of some downstream tasks in our experiments may be making it easier for the models to perform zero-shot extrapolation. We perform a 0-shot extrapolation test with different extrapolation lengths with the Masked Language Modeling (MLM) loss as our metric during validation for a Hawk model being trained on 12kbp sequences. The results in Figure~\ref{fig:val_loss_comp} also underscore a key theme in our paper: the capacity of state-space models (SSMs) to extrapolate effectively to longer contexts with minimal performance degradation. By evaluating pure MLM loss—independent of downstream tasks—we isolate how well the model handles missing-token predictions at varying lengths. The near-uniform loss profiles across different sequence lengths support the broader findings of the paper, namely that SSM-based architectures learn high quality internal representation that allow them to retain their predictive accuracy and generalization properties even beyond the context lengths they were originally trained on.

\section{Discussion and Conclusion}

Our evaluation of State-Space Models (SSMs) for long-range genomic modeling demonstrates that these architectures learn high-quality representations that are both biologically meaningful and computationally scalable. Across multiple downstream tasks, SSMs not only match transformer performance but also excel in zero-shot extrapolation—extending from a 12\,kbp training context to sequences up to 120\,kbp and even 1\,Mbp without additional fine-tuning. This behavior aligns with our goal of capturing the genome's hierarchical regulation, preserving both fine-grained nucleotide details and long-range regulatory interactions.

\subsection{Expanded Analysis of Key Findings}

Our experiments reveal several critical insights:

\textbf{Performance on Downstream Genomic Tasks:}  
SSM-based models demonstrate competitive performance on standard tasks such as Bulk RNA prediction, VEP ClinVar, and Histone Marks detection. In several cases, models like Caduceus even outperform the transformer baseline (NTv2), underscoring the capability of SSMs to capture intricate genomic patterns despite their reduced parameter count and improved computational efficiency.

\textbf{Zero-shot Extrapolation:}  
A standout result is the ability of SSMs to generalize to input sequences significantly longer than those encountered during training. Figures~\ref{fig:extrapolation_ntv2_50m}, ~\ref{fig:zero_shot_extrapolation}, and \ref{fig:Hawk-1M} illustrate that the recurrent state propagation inherent in SSMs allows for 10-100 $\times$ increase in sequence lengths beyond the training context without any considerable decline in performance. In fact, the trend lines suggest they can process even longer sequences without suffering performance declines. This zero-shot extrapolation is especially important for genomics, where regulatory interactions can span hundreds of kilobases or more. The consistent performance across varied sequence lengths suggests that SSMs capture distributed dependencies in a way that aligns with the biological reality of long-range interactions.

\textbf{Consistency of Pretraining Loss:}  
Our analysis of Masked Language Modeling (MLM) loss (Figure~\ref{fig:val_loss_comp}) shows nearly uniform loss values across a wide range of sequence lengths. This consistency reinforces the idea that SSMs build robust internal representations capable of handling extended contexts. The stable loss behavior provides additional evidence that these models are not simply memorizing local patterns but are instead learning scalable representations that remain effective when applied to ultralong genomic sequences.

\textbf{Biological Relevance and Representation Quality:}  
The observed performance, combined with the ability to extrapolate effectively, implies that SSMs capture the intrinsic organization of genomic data. By preserving both the nucleotide-level details and the extended regulatory networks, these models generate representations that mirror the true hierarchical nature of the genome. This fidelity is critical for modeling biological phenomena such as enhancer–promoter interactions and chromatin remodeling, which are essential for understanding gene regulation, disease mechanisms, and evolutionary processes.

\subsection{Implications and Future Directions}

The results presented in this work highlight the potential of SSM-based architectures as scalable alternatives for comprehensive genomic analysis. The demonstrated ability to process ultralong sequences on a single GPU not only makes these models practical for large-scale studies but also opens the door for integrated analyses of entire genomic regions in one pass. Future research should focus on:
\begin{itemize}
\item \textbf{Architectural Refinement.} Further enhancing model designs to improve task-specific performance while retaining zero-shot extrapolation. Investigating hybrid models that combine the benefits of state-space and attention-based mechanisms may yield additional gains.
\item\textbf{Expanded Evaluation.} Applying these models to a broader array of genomic benchmarks to validate their versatility across different biological contexts, thereby solidifying their utility in diverse genomic applications.
\item \textbf{Improving utilization of extended context.} While we show SSMs can reliably extrapolate to longer contexts, we don't see a significant improvement in downstream tasks despite the added contextual information. Analyzing this could give us insights into utilizing longer contexts more thoroughly, thereby improving their performance on downstream tasks.
\end{itemize}
\subsection{Concluding Remarks}

In summary, our integrated results demonstrate that SSM-based models not only achieve competitive performance on traditional genomic tasks but also offer significant advantages in capturing long-range dependencies through zero-shot extrapolation and consistent pretraining behavior. By faithfully representing both local and global genomic features, these models provide a computational framework that aligns with the inherent complexity of biological systems. As we refine these architectures and extend their application, we are poised to drive new discoveries in gene regulation, disease mechanisms, and evolution—ultimately capturing a more meaningful representation of life.

\newpage

\textbf{Meaningfulness Statement.} Meaningful representations of life capture the intrinsic organization and multi-scale interactions of biological systems. In genomics, this means preserving the fine-grained nucleotide details and the long-range regulatory networks that govern gene expression, enabling broad applicability across biological tasks, i.e. generalization. In our work, we show that SSMs create such representations, and show utility and generalization across diverse tasks and input distributions (sequence lengths).

\bibliography{iclr2025_conference}
\bibliographystyle{iclr2025_conference}

\newpage

\appendix
\section{Appendix}

\begin{figure}[H]
    \centering
    \begin{subfigure}[b]{0.45\textwidth}
        \centering
        \includegraphics[width=\textwidth]{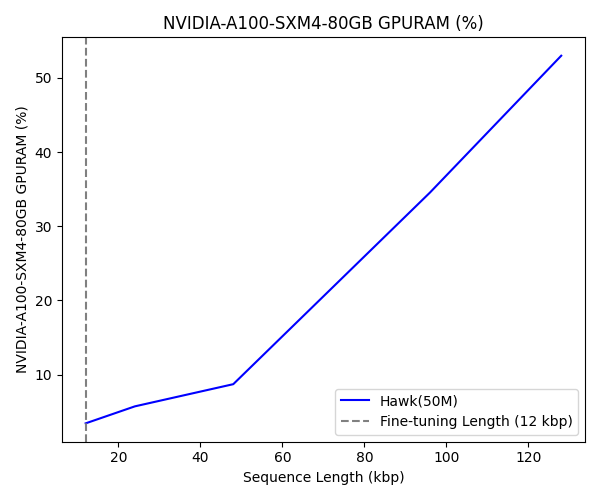}
        \caption{GPU Memory Usage vs.\ Sequence Length}
        \label{fig:griffin_memory}
    \end{subfigure}
    \hfill
    \begin{subfigure}[b]{0.45\textwidth}
        \centering
        \includegraphics[width=\textwidth]{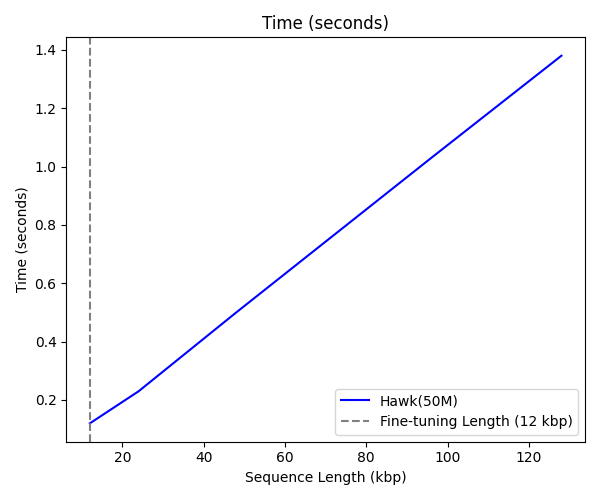}
        \caption{Inference Time vs.\ Sequence Length}
        \label{fig:griffin_time}
    \end{subfigure}
    \label{fig:griffin_perf}
\end{figure}

An equally important advantage of SSM-based architectures is their computational efficiency. Figures~\ref{fig:griffin_memory} and~\ref{fig:griffin_time} report GPU memory usage and inference time of Hawk as functions of input sequence length. The linear scaling observed for both metrics confirms that SSMs can process ultralong sequences with a computational cost that increases only linearly with sequence length—an improvement over the quadratic scaling typical of transformer-based models. This efficiency not only enables the processing of million-token sequences on a single GPU but also highlights the practical feasibility of applying SSMs to large-scale genomic datasets. 
Our Hawk model is 12 LRU blocks—each featuring 6 attention heads, a hidden width of 576, and a feed-forward expansion factor of 3 (yielding 1728) and LRU block width of 768. During pretraining we used a warmup-cosine decay from 1e-8 up to 1e-4 learning rate, and applied gradient accumulation to reach an effective batch size of 96. 

\section{Comparison of Positional Encodings in Transformers and State-Space Models for Long Genomic Sequences}

Transformers explicitly encode positional information through mechanisms such as sinusoidal embeddings \citep{vaswani2023attentionneed}, Rotary Position Embeddings (RoPE) \citep{su2024roformer}, YaRN \citep{peng2023yarnefficientcontextwindow}, and SelfExtend \citep{jin2024llmmaybelonglmselfextend}. Each of these methods encounters inherent limitations when extrapolating far beyond their training sequence length, critically limiting their effectiveness in ultralong genomic contexts. Specifically, sinusoidal embeddings, which rely on fixed periodic functions, become ambiguous at positions significantly longer than the training range, resulting in diminished predictive accuracy. RoPE embeddings attempt to improve this by using rotational transformations to encode relative positions; however, at large genomic distances, these rotations become increasingly misaligned, undermining accuracy in modeling long-range interactions like enhancer-promoter dynamics.

Further enhancements, including YaRN and SelfExtend, partially address this limitation through positional interpolation or grouping. However, these techniques either require supplementary fine-tuning or lose essential positional granularity, which is vital for accurately modeling subtle yet biologically meaningful genomic relationships.

In contrast, State-Space Models (SSMs) implicitly capture positional information through continuous hidden-state updates, providing inherent scalability and robust positional encoding without recalibration. This implicit encoding maintains accuracy across extreme sequence lengths, effectively modeling genomic interactions at high resolution and addressing fundamental extrapolation limitations faced by transformers. We hypothesize that this intrinsic mechanism is one of the primary factors contributing to the strong performance of SSMs in zero-shot long-range extrapolation observed in our work.

\bibliographystyle{iclr2025_conference}
\bibliography{iclr2025_conference}

\end{document}